\documentclass[prd,twocolumn,floatfix,preprintnumbers]{revtex4}

\usepackage{graphicx}
\usepackage{graphicx,epsfig}
\usepackage{bm}
\usepackage{latexsym,amssymb,amsmath,float}

\newcommand {\ga} {\ {\raise-.5ex\hbox{$\buildrel>\over\sim$}}\ }
\newcommand {\la} {\ {\raise-.5ex\hbox{$\buildrel<\over\sim$}}\ }

\def\be{\begin{equation}}
\def\ee{\end{equation}}
\def\ba{\begin{eqnarray}}
\def\ea{\end{eqnarray}}

\begin{document}

\title{Inflection Point Quintessence}
\author{Hui-Yiing Chang and Robert J. Scherrer}
\affiliation{Department of Physics and Astronomy, Vanderbilt University,
Nashville, TN  ~~37235}

\begin{abstract}
We examine models in which the accelerated expansion of the universe is driven
by a scalar field rolling near an inflection point in the potential.
For the simplest such models, in which the potential is of
the form $V(\phi) = V_0 + V_3 (\phi-\phi_0)^3$, the scalar field can either evolve toward $\phi = \phi_0$
at late times, yielding
an asymptotic de Sitter expansion, or it can transition through the
inflection point, producing a transient period of acceleration.  We
determine the parameter ranges which produce each of these two possibilities
and also map out the region in parameter space for which the equation of state
of the scalar field is close to $-1$ at all times up to the present,
mimicking $\Lambda$CDM.
We show that the
latter can be consistent with either eternal or transient acceleration.  More
complicated inflection point models are also investigated.
\end{abstract}

\maketitle

\section{Introduction}

Cosmological data \cite{Knop,Riess,union08,perivol,hicken,Hinshaw,Ade}
indicate that roughly
70\% of the energy density in the
universe is in the form of a negative-pressure component,
called dark energy, with roughly 30\% in the form of nonrelativistic matter (including both baryons
and dark matter).
The dark energy component can be parametrized by its equation of state parameter, $w$,
defined as the ratio of the dark energy pressure to its density:
\be
\label{w}
w=p/\rho.
\ee
A cosmological constant, $\Lambda$, corresponds to the case $w = -1$ and $\rho = constant$.

While a model with a cosmological constant and cold dark matter ($\Lambda$CDM) is consistent
with current observations,
there are many realistic models of the Universe that have a dynamical equation
of state.
For example, one can consider quintessence models, with a time-dependent scalar field, $\phi$,
having potential $V(\phi)$
\cite{RatraPeebles,CaldwellDaveSteinhardt,LiddleScherrer,SteinhardtWangZlatev}.
(See Ref. \cite{Copeland} for a review). 

In order to produce a present-day value of $w$ close to $-1$, we require $p \approx - \rho$,
so that $\dot \phi^2 \ll V(\phi)$ at present. One way to achieve this is
for $\phi$ to be located in a very flat portion of the potential, so that
\be
\label{flat}
\left(\frac{1}{V} \frac{dV}{d\phi}\right)^2 \ll 1.
\ee
Several previous papers have investigated such models
in which equation (\ref{flat}) is satisfied when the potential is close to linear
\cite{ScherrerSen1,ScherrerDutta1} or close
to a local maximum \cite {ds1} or minimum \cite{ds3}.

Here we examine the next higher-order extension of this idea:  quintessence with a
scalar field evolving near an inflection point of the potential.  Scalar field
models with an inflection point in the potential 
have been investigated previously
in connection with inflation 
\cite{Allahverdi1,Allahverdi2,Allahverdi3,Baumann,Panda,Itzhaki1,Krause,Allahverdi4,Itzhaki2,Enqvist,Hotchkiss,
Chatterjee,Downes1,Downes2,
Choudhury} and
have been dubbed ``inflection point inflation."  The major
difference between these inflation models and the inflection point quintessence models
we examine here is that inflation takes place in a scalar-field-dominated universe,
while for the case of dark energy, we are interested in the evolution of the quintessence field
at low redshift, when the Friedman equation must include both the scalar field and nonrelativistic matter.
Thus, results from inflection point models for inflation will not necessarily carry over
into inflection point quintessence.

In investigating the evolution of inflection point quintessence, there are two important
questions to address.  The first is whether the scalar field rolls slowly enough near
the inflection point to generate the observed value of $w$ near $-1$, for consistency with
the observations.  The second issue is whether $\phi$ evolves to a constant value
at the inflection point, generating a model for which $w \rightarrow -1$ asymptotically, and
yielding a model essentially indistinguishable
from a cosmological constant, or whether $\phi$ rolls
through the inflection point, so that $w$ deviates away from $-1$ eventually.  The latter
possibility would produce a transient stage of acceleration, rather than an asymptotic de Sitter
evolution.  This is of interest because an eternally accelerating
universe presents a problem for string theory, inasmuch as the S-matrix in this case is ill-defined
\cite{Hellerman,Fischler}.  Consequently, some effort has gone into the development of models in which the
observed acceleration is a transient phenomenon \cite{Barrow,Cline,Cardenas,Sahni,Blais,Bilic,GCG,Carvalho,
Bento,Cui}.  Our model represents another example of this
sort of transient acceleration for the case in which $\phi$ evolves through the inflection point.

In the next section, we present the general models under discussion.  Unlike the linear and quadratic potentials
examined in Refs. \cite{ScherrerSen1,ScherrerDutta1,ds1,ds3}, the simplest version of
inflection point quintessence, with a cubic term in the potential, does not yield a simple analytic expression for the evolution,
so we solve it numerically in Sec. IIa and determine
the regions in parameter space for which the model produces transient or eternal acceleration.
We also determine the range of parameters for which $w \approx -1$ at all times up to the present.
In Sec. IIb, we examine other inflection point models,
demonstrating that some of these do have analytic descriptions for their behavior.
Our conclusions are summarized in Sec. III.

\section{The inflection point quintessence model}

We assume that the dark energy is given by a minimally coupled
scalar field, with equation of motion
\be
\label{phievol0}
\ddot{\phi} + 3H \dot{\phi} + \frac{dV}{d\phi} = 0,
\ee
where the dot indicates the derivative with respect to time, and $H$ is the Hubble parameter,
given by
\be
\label{Hubble}
H^2 = {\left ( \frac{\dot{a}}{a} \right )}^2 = \frac{\rho_\phi + \rho_M}{3},
\ee
where we assume a flat Universe and consider only times sufficiently late
that the expansion is dominated by matter and dark energy.
(We take $\hbar = c = 8 \pi G=1$ throughout).
In Eq. (\ref{Hubble}),
the scalar field energy density is
\be
\rho_\phi={\frac{1}{2}}{\dot{\phi}}^2+V(\phi),
\ee
and the matter energy density is
\be
\rho_M = \rho_{M0} a^{-3}.
\ee
The scalar field pressure is
\be
p_\phi={\frac{1}{2}}{\dot{\phi}}^2-V(\phi),
\ee
and the equation of state parameter, $w$, is given by Eq. (\ref{w}).

We will consider the general case of potentials with an inflection point in the potential.
The simplest example
of such a model is a potential of the form
\be
\label{ip3}
V(\phi) = V_0 + V_3 (\phi- \phi_0)^3,
\ee
which has an inflection point with $dV/d\phi = 0$ at $\phi= \phi_0$.  This
is the potential that we will investigate in the next section.  It is not
only the simplest inflection point quintessence model, but as we shall see,
it also produces
some of the most interesting behavior.

The potential in Eq. (\ref{ip3}) can be generalized in several ways.  For instance,
one can add linear and quadratic terms to obtain
\be
\label{ip3general}
V(\phi) = V_0 + V_1 (\phi-\phi_0) + V_2 (\phi-\phi_0)^2 + V_3 (\phi - \phi_0)^3.
\ee
It is also possible to consider more general inflection points produced by models with other
powers of $\phi$ in the potential, i.e., potentials
of the form 
\be
\label{ipn}
V(\phi) = V_0 + V_n (\phi-\phi_0)^n.
\ee
We will examine the models given by Eqs. (\ref{ip3general})-(\ref{ipn}) in Sec.
IIb.

Since we are interested in the behavior of the scalar field near the inflection point,
we will take Eqs. (\ref{ip3})-(\ref{ipn}) to refer only to the behavior
of $V(\phi)$ in the region near the inflection point and make
no assumptions about what the
rest of the potential
looks like.  Thus, our results will be more general than if we had
assumed that these were the exact forms for
the potential for all values of $\phi$.  Furthermore, the fact that these potentials
are not bounded from below as $\phi \rightarrow -\infty$ is not pathological,
since we do not assume that Eqs. (\ref{ip3})-(\ref{ipn}) apply in this limit.

\subsection{The cubic inflection point model}

Consider first the potential given by Eq. (\ref{ip3}).  Equation (\ref{phievol0})
becomes
\be
\label{phievol1}
\ddot{\phi} + 3H \dot{\phi} + 3V_3(\phi-\phi_0)^2 = 0.
\ee
The evolution of $\phi$ is specified by four parameters, namely
the initial values of $\phi$ and $\dot \phi$, and the values of $V_0$ and
$V_3$.  However, some simplifications are possible.

We first note that the value of $\phi_0$ has no effect on any physically
observable quantities, so we can redefine the field to take $\phi_0 = 0$,
and equation (\ref{phievol1}) becomes
\be
\label{phievol}
\ddot{\phi} + 3H \dot{\phi} + 3V_3\phi^2 = 0,
\ee
with $H$ given by
\be
\label{H2}
H^2 = \frac{1}{3} \left(
\rho_{M0}a^{-3} + {\frac{1}{2}}{\dot{\phi}}^2+ V_0 + V_3\phi^3\right).
\ee
In order for the models examined here to be consistent with current
observations, they must closely resemble $\Lambda$CDM, which is possible
only if $V_0$ corresponds to the observed present-day dark energy density.
Having fixed $V_0$, we can completely specify the models by the value of $V_3/V_0$.  We will further assume for simplicity
that $\dot \phi_i = 0$, i.e., the field is initially at rest.  For many models
of interest the damping term in Eq. (\ref{phievol0}) will tend to drive
$\dot \phi$ to $0$ at early times, giving the initial condition we consider
here.

Hence, we are left with a model that is completely specified by $V_3/V_0$
and by the initial value of the scalar field, $\phi_i$.  We have numerically
integrated Eqs. (\ref{phievol}) and (\ref{H2}) to determine the behavior of
$\phi(t)$ as a function of $V_3/V_0$ and $\phi_i$.  We find two distinct
possible behaviors for $\phi$:  the field can either evolve past the inflection
point at $\phi=0$, or else it can evolve smoothly to $\phi=0$ as $t \rightarrow \infty$.
These two different types of behavior are shown in Fig. 1.  The two $\phi(t)$ trajectories
in this figure both have $V_3/V_0 = 1$, but slightly different initial values of $\phi$, resulting
in nearly identical evolution until the field approaches the inflection point,
where
the trajectories diverge to give very different asymptotic behaviors.

We find that for fixed $\phi_i$, a sufficiently large value of $V_3/V_0$
causes the scalar field to evolve through the inflection point, while
for smaller values of $V_3/V_0$ the field evolves asymptotically to $\phi=0$
as $t \rightarrow \infty$.  This is illustrated in Fig. 2.  The region above
the solid (black) curve gives a field that evolves through the inflection point,
while the region below this curve has $\phi \rightarrow 0$ as $t \rightarrow
\infty$.  Note, however, that for sufficiently small values of $V_3/V_0$,
the field never transitions through the inflection point for any value of
$\phi_i$ so the solid curve becomes a horizontal line for large $\phi_i$.
We can define a critical value, $(V_3/V_0)_c$, below which evolution through the
inflection point becomes impossible.  Our numerical results indicate that
$0.77 < (V_3/V_0)_c < 0.78$.

\begin{figure}[t!]
\centerline{\epsfxsize=3.8truein\epsffile{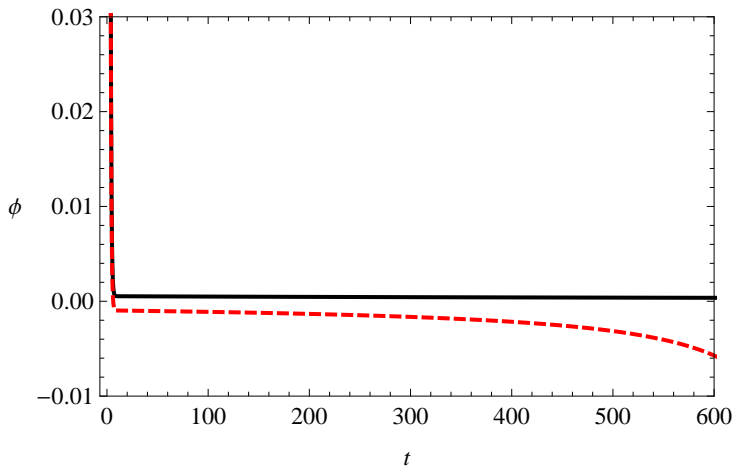}}
\caption{The evolution of the scalar field $\phi$ as a function of time $t$
for the potential $V(\phi) = V_0 + V_3 \phi^3$, with $V_3/V_0 = 1$.
Black (solid) curve is for $\phi_i = 1.76$;
red (dashed) curve is for $\phi_i = 1.78$.}
\end{figure}

Note that a similar study was undertaken by Itzhaki and Kovetz
\cite{Itzhaki2}, who explored the asymptotic evolution of the scalar field
in inflection point inflation.  Their study differs from ours in that we
include nonrelativistic matter, which alters the evolution of $H$ in equation
(\ref{phievol}).  However, we find that matter is subdominant as
$\phi \rightarrow 0$ for the model parameters lying along
the transition regime defined by the black curve
in Fig. 2.  Hence, we would expect our results to agree with Ref.
\cite{Itzhaki2}
with regard to the existence of a critical value of $V_3/V_0$ below which
the field can never cross the inflection point, and Itzhaki and Kovetz do,
indeed, observe such behavior.  They find $(V_3/V_0)_c = 0.7744$, in agreement
with our results for the quintessence model.

The results displayed in Figs. 1-2 show that the inflection point quintessence
model can lead to two very different future evolutionary paths for the universe.
When $\phi$ asymptotically goes to zero, we are left with a model
essentially identical to $\Lambda$CDM, with a de Sitter evolution.  However,
the second possibility, in which $\phi$ evolves through the inflection point, yields
a model in which the accelerated expansion of the universe is a transient phenomenon.
As noted earlier, we make no assumptions about $V(\phi)$ far from the inflection
point, so the future evolution of the universe in this case will depend on the particular
form for $V(\phi)$ with $\phi < 0$.

Clearly the first possibility can be made consistent with the observations, since the
current observational data is well-fit by $\Lambda$CDM.  A more interesting question is
whether the models with transient acceleration, in which $\phi$ passes through the
inflection point, can be made consistent with the observations.  Here we demonstrate
a stronger result:  models with transient acceleration can mimic
$\Lambda$CDM at all times up to the present.  In our model, $w=-1$ initially,
since the field begins with $\dot\phi = 0$.  As the field begins to roll
down the potential, $w$ increases away from $-1$.  In Fig. 2 we have mapped out the regions in parameter space
for which $-1 < w < -0.95$ at all times up to the present (which we take to correspond
to $\Omega_\phi = 0.7$); this is the region below the green dashed curve.  The region below
the red dotted curve defines the set of parameters for which $-1 < w < -0.9$ at all times
up to the present.  Thus, the region between the black curve and the green
curve is essentially indistinguishable from $\Lambda$CDM on the basis of current
observations, and yet it results in an evolution in which the
current accelerated phase of the expansion will eventually come to an end.

\subsection{Other inflection point models}

Although the simplest inflection point potential is the cubic potential
considered in
the previous section, more general models of the form
\begin{eqnarray}
\label{Vn}
V(\phi) &=& V_0 + V_n \phi^n,~~~{\rm n~odd},\nonumber\\
V(\phi) &=& V_0 + V_n {\rm sgn}(\phi) \phi^n,~~~{\rm n~even},
\end{eqnarray}
are also possible.
Models of this sort were discussed briefly in Ref. \cite{Itzhaki2} in the
context of inflation.

\begin{figure}[t!]
\centerline{\epsfxsize=3.8truein\epsffile{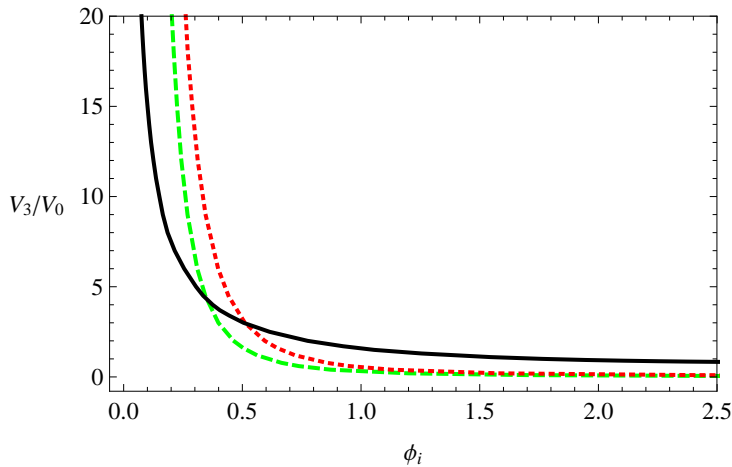}}
\caption{For the scalar field potential $V(\phi) = V_0 + V_3 \phi^3$, the curves
divide regions with different behaviors for $\phi$ as a function of the initial
value of the field, $\phi_i$, and the ratio of $V_3$ to $V_0$.  Above and to the
right of the black
(solid) curve, the field evolves through the inflection point at $\phi=0$, while
below and to the left of this curve, $\phi \rightarrow 0$
as $t \rightarrow \infty$.  The regions below the green (dashed) and dotted
(red) curves closely mimic $\Lambda$CDM.  The region below the green curve
has an equation of state parameter for $\phi$ satisfying $-1 < w < -0.95$ at all times up
to the present, while the region below the red curve corresponds to
$-1 < w < -0.9$.}
\end{figure}

Consider first the case $n=2$.  The evolution of a quintessence field
in a potential of the form
\be
\label{harmonic}
V = V_0 + V_2 \phi^2,
\ee
was analyzed in Ref.
\cite{ds3}, based on the results of Ref. \cite{ds1}, for a universe
containing both matter and a scalar field, in the limit where
\be
\label{condition}
\frac{1}{V}\frac{dV}{d\phi} \ll 1.
\ee
In this limit, two behaviors are possible, based on the value of
$(1/V)(d^2 V/d\phi^2)$ at the minimum of the potential.  When
$(1/V)(d^2 V/d\phi^2) < 3/4$ at the minimum of the potential, the scalar field
asymptotically approaches $0$, while for $(1/V)(d^2 V/d\phi^2) > 3/4$, the field
oscillates around the minimum \cite{ds3}.

The $n=2$ case of Eq. (\ref{Vn}) is identical to Eq. (\ref{harmonic}) for $\phi
> 0$, so the results of Ref. \cite{ds3} carry over directly to the inflection
point case:  for $(1/V)(d^2 V/d\phi^2) < 3/4$ the scalar field will never cross
the inflection point, while for $(1/V)(d^2 V/d\phi^2) > 3/4$ the field will evolve
across the inflection point and the accelerated expansion will be transient.

In terms of the parameters $V_0$ and $V_2$ in Eq. (\ref{Vn}), the condition
for evolution through the inflection point becomes
\be
\label{2tunnel}
\frac{V_2}{V_0} > \frac{3}{8}.
\ee
Conversely, when Eq. (\ref{2tunnel}) is not satisfied, the scalar field evolves
to $\phi =0$ asymptotically.
The ``slow-roll" condition for the results of Ref. \cite{ds3}
to be valid (Eq. \ref{condition}) will be satisfied for
\be
\label{valid}
\frac{2 V_2 \phi}{V_0} \ll 1.
\ee
Note, however, that as $\phi \rightarrow 0$, Eq. (\ref{valid}) is always eventually satisfied,
so Eq. (\ref{2tunnel}) provides the correct condition for
evolution through the inflection point for arbitrary initial values of $\phi$;
we have verified this result numerically.

Thus, the evolution of $\phi$ for the $n=2$ case of Eq. (\ref{Vn}) qualitatively
resembles the results shown in Fig. 2 for the cubic inflection point potential
when the latter
has large $\phi_i$, albeit with a different critical value for $V_n/V_0$.
Thus, while the $n=2$ case is somewhat unnatural, it does provide insight into
the qualitative behavior of the more interesting $n=3$ case.

However, the behavior of these two models is quite different for small $\phi_i$.
In this case, transition through the inflection point for
$n=3$ requires increasingly large values of $V_3/V_0$ as $\phi_i \rightarrow 0$,
while for $n=2$ the critical value for $V_2/V_0$ remains constant for all
$\phi_i$.

Now consider larger values of $n$.   The evolution of scalar field potentials
of the form
\be
V(\phi) = V_n\phi^n
\ee
for the case of a matter-dominated expansion was
previously examined in Ref. \cite{LiddleScherrer}:
for $n > 6$, the field evolves smoothly to $\phi \rightarrow 0$ as
$t \rightarrow \infty$ (see also the discussion in Ref. \cite{Oscil}).
In the case considered here, we have an additional
contribution to $H$ in Eq. (\ref{phievol0}):  the contribution
of the scalar field energy density, which now also includes
the additional constant
term, $V_0$, in the scalar field potential.
However, this additional contribution to the friction term in Eq.
(\ref{phievol0}) can only serve to decrease $\dot \phi$ as the scalar field
rolls toward the inflection point, making it more difficult for the field to
reach $\phi=0$.  Thus, we can conclude from the results
of Ref. \cite{LiddleScherrer} that inflection point potentials
of the form given in Eq. (\ref{Vn}) with $n > 6$ never transition through
the inflection point.

The cases $n=4,5,6$ are neither as interesting as $n=3$ nor as amenable to
analytic solution as $n=2$ or $n>6$, so we will not discuss them in detail here.
 However, numerical integration indicates that, like the $n=3$ case, they can
 yield either evolution of the scalar field through the inflection point, or
 attraction to the inflection point, depending on the model parameters.

The simple cubic inflection point model given by Eq. (\ref{ip3}) can also be
generalized by adding linear and quadratic terms as in Eq. (\ref{ip3general}).
Note that the quadratic term can be eliminated by a suitable translation
of $\phi$, and we can set the corresponding (new) value of $\phi_0$ to zero as in
the previous section, to yield
\be
\label{fullmodel}
V(\phi) = V_0 + V_1 \phi + V_3 \phi^3.
\ee

The evolution of $\phi$ will then depend on the sign of $V_1$.
For
$V_1 > 0$, the potential has no local minima, and the field will always
transition through the inflection point at $\phi=0$.  However, for
sufficiently small $V_1$, one can still have a model arbitrarily close
to $\Lambda$CDM.  Thus, in this variant of inflection point quintessence,
the accelerated expansion of the universe is always a transient phenomenon.

On the other hand, if $V_1 < 0$, the potential develops a local minimum at
$\phi = \sqrt{-V_1/3V_3}$ and a local maximum at $\phi = -\sqrt{-V_1/3V_3}$.
Depending on the parameter values and initial conditions for $\phi$, it is
possible for the field to transition through the local maximum, so that the
accelerated expansion of the Universe is transient, or to get trapped in the local minimum, producing
eternal acceleration.  This behavior resembles the evolution of the scalar
field in the Albrecht-Skordis model \cite{AS}, in which the
potential is given by the product of an exponential and a polynomial,
producing local minima in $V(\phi)$.
In the Albrecht-Skordis model, the accelerated
expansion of the universe can be either permanent or transient, depending
on the model parameters \cite{Barrow}.

\section{Conclusions}

Inflection point quintessence represents an interesting new model for the dark
energy that drives the accelerated expansion of the universe.
Even the simplest form of this model, with the
potential given by Eq. (\ref{ip3}) and the scalar field
initially at rest, displays a variety of intriguing
behaviors.  For large initial values of $\phi$,
the asymptotic
behavior of $\phi$ becomes independent of $\phi_i$ and depends only on $V_3/V_0$,
while for small $\phi_i$, the behavior depends
on both $V_3/V_0$ and $\phi_i$.  In either case, it is possible to have asymptotic
evolution for which $\phi \rightarrow 0$, and the Universe undergoes eternal
de Sitter expansion, or, conversely, for $\phi$ to transition through the inflection point, leading
to transient acceleration.

It is interesting to note that this potential can yield an attractor at $\phi = 0$
despite the fact that $V(\phi)$ has an inflection point, rather than a local
minimum, at $\phi=0$.  On the other hand,
it is possible to construct models
very close to $\Lambda$CDM today which nonetheless evolve away from accelerated
expansion in the future.

Inflection point quintessence shows, within the context of a very simple model, that
current data may never be sufficient to determine
whether the universe will accelerate forever or simply pass through a transient
period of acceleration.  While we have not explored in similar detail the more general
inflection point models
given by Eq. (\ref{fullmodel}), these models, too,
can give rise to either eternal de Sitter expansion or transient acceleration, but in this
case the asymptotic behavior depends strongly on the sign of the linear term.

\section{Acknowledgments}
R.J.S. was supported in part by the Department of Energy
(DE-FG05-85ER40226).

\end{document}